\documentclass[sigconf]{acmart}
\usepackage{booktabs} 
\usepackage[]{todonotes}
\usepackage{listings} 
\usepackage{newfloat}
\usepackage[inline]{enumitem}
\usepackage{caption}
\usepackage{enumitem}
\usepackage{color}
\usepackage{amsmath}
\usepackage{amssymb}
\usepackage[T1]{fontenc}
\usepackage[utf8]{inputenc}
\usepackage{hyperref}
\usepackage{tabularx}
\usepackage{bbding}
\usepackage[edges]{forest}
\usetikzlibrary{arrows.meta}
\usepackage{tikz-qtree}

\DeclareFloatingEnvironment[
fileext=ltt,
listname={List of Listings},
name=Listing
]{listing}

\newcommand{\stride}{\textsc{Stride}}
\newcommand{\estride}{\textsc{eStride}}
\newenvironment{myquote}[1]%
{\list{}{\leftmargin=#1\rightmargin=#1}\item[]}%
{\endlist}

\begin{document}
\title{Finding Security Threats That Matter: An Industrial Case Study}

\author{Katja Tuma $\dagger$, Christian Sandberg, Urban Thorsson, Mathias Widman, Riccardo Scandariato $\dagger$}
\affiliation{
\textit{$\dagger$ University of Gothenburg} \textit{and Chalmers University of Technology} \\
}
\affiliation{ 
	\institution{\textit{AB Volvo} \\ \textit{Gothenburg, Sweden}}
}
\email{katja.tuma@cse.gu.se, christian.sandberg, urban.thorsson, mathias.widman@volvo.com, riccardo.scandariato@cse.gu.se}

\begin{abstract}
Recent trends in the software engineering (i.e., Agile, DevOps) have shortened the development life-cycle limiting resources spent on security analysis of software designs.
In this context, architecture models are (often manually) analyzed for potential security threats.
Risk-last threat analysis suggests identifying all security threats before prioritizing them.
In contrast, risk-first threat analysis suggests identifying the risks before the threats, by-passing threat prioritization.
This seems promising for organizations where developing speed is of great importance.
Yet, little empirical evidence exists about the effect of sacrificing systematicity for high-priority threats on the performance and execution of threat analysis.
To this aim, we conduct a case study with industrial experts from the automotive domain, where we empirically compare a risk-first technique to a risk-last technique.
In this study, we consciously trade the amount of participants for a more realistic simulation of threat analysis sessions in practice.
This allows us to closely observe industrial experts and gain deep insights into the industrial practice.
This work contributes with: 
(i) a quantitative comparison of performance, 
(ii) a quantitative and qualitative comparison of execution, and
(iii) a comparative discussion of the two techniques.
We find no differences in the productivity and timeliness of discovering high-priority security threats.
Yet, we find differences in analysis execution.
In particular, participants using the risk-first technique found twice as many high-priority threats, developed detailed attack scenarios, and discussed threat feasibility in detail.
On the other hand, participants using the risk-last technique found more medium and low-priority threats and finished early.
\end{abstract}

\keywords{Threat analysis, Risk analysis, STRIDE, Case study, Empirical software engineering}

\maketitle


\section{Introduction}
\label{sec:intro}
\looseness=-1
Security threats to software systems are a great concern in many organizations.
A recent survey by Ernst \& Young~\cite{EYglobalsurvey} shows that $53\%$ of organizations have increased their cybersecurity budget in the last year, and further $65\%$ of organizations foresee a further increase in the coming year.
Increased investments in security come as a response to recent changes in legislation regarding data privacy (GDPR) and upcoming security standards (e.g., ISO 21434). 
Studies have shown \cite{cavusoglu2015institutional,moore2016identifying} that compliance is a key factor for investing in security.
For example, Cavusoglu et al.~\cite{cavusoglu2015institutional} investigate the influence of institutional pressures on security investment. 
The authors find that aside from normative pressure (manifested from the state-of-practice), coercive pressure (manifested by perceived pressure from business partners and pressure from regulations) is one of the key factors impacting investments in security resources.

\looseness=-1
Security-by-design techniques help planning for security early in the design phase where little is known about the system \cite{mcgraw2006software}.
In this context, architectural design models are analyzed for security \cite{howard2006security}. 
Threat analysis techniques facilitate a systematic analysis of the attacker's profile, vis-a-vis the assets of value to the organization \cite{tuma2018threat}.
For instance, STRIDE is a well-known threat analysis technique that is also used in the automotive domain. 
This technique has the tendency to produce a high volume of threats~\cite{tuma2018two,scandariato2015descriptive}.
Many threats are later-on discarded due to a low risk value (a combination of impact and likelihood). 
In this respect, STRIDE is a risk-last approach, where risk is considered only \textit{after} the threats have been analyzed. 
Note that this way of working is inefficient, as significant time and effort is spent on discussing unimportant threats. 

Many organizations have undergone cultural change and adopted to agile development practices, cross-functional teams, continuous integration (CI), development and operations (DevOps), and the like \cite{dybaa2008empirical}.
Recent trends in software engineering and development have the ambition to shorten the software development life-cycle. 
Release cycles are often shortened to days, or even a few hours.
Yet, analysis of software design for security often requires security expertise and is time-consuming.
Therefore, in this context, software design is often neither documented nor analyzed for potential security issues.
Further, practices such as prototyping and refactoring require repeating entire analysis sessions to determine the impact of new design decisions on security.
Such practices would require more \textit{efficient} security-by-design approaches.

\looseness=-1
The recently proposed eSTRIDE \cite{tuma2017towards} is a risk-first threat analysis approach.
This approach is accompanied by an extended model notation (eDFD).
The model is extended with information about important assets, security assumptions, and existing security mechanisms.
These extensions are leveraged during diagram analysis to focus on critical parts of the architecture, by-passing threat prioritization all together.
This seems promising for organizations with development speed requirements.
Yet, little empirical evidence exists about the effect of sacrificing systematicity for high-priority threats on threat analysis performance and execution.

The purpose of this study is to gather evidence about the similarities and differences between a risk-first and a risk-last threat analysis technique in an industrial setting.
To this aim, we conduct a case study with industrial participants from the automotive domain.
The case study is designed to simulate the industrial practice as much as possible. 
For instance, the industrial case is selected by co-authors, who participate in threat analysis sessions of real cases on a daily basis.
Further, we simulate time-constraints that are present in practice by limiting the amount of available time to our participants (in total 6 hours for the complete analysis session). 
Finally, this study observes two teams (7 participants in total) to compare the analysis performance and execution.
This design choice was made explicitly because it allowed us to better monitor the teams (in total, 9 hours of transcribed recordings) and gain deep insights about the analysis execution.

The contributions of this work are three-fold: 
(i) a quantitative comparison of technique performance,
(ii) a quantitative and qualitative comparison of technique execution, and
(iii) a comparative discussion of the benefits and shortcomings of two techniques, in a realistic setting.
This study shows no significant differences in productivity and timeliness of discovering high-priority security threats.
But, we find differences in analysis execution.
Specifically, participants using the risk-first technique found twice as many high-priority threats, developed detailed attack scenarios, and discussed threat feasibility in greater detail.
On the other hand, participants using the risk-last technique found more medium and low-priority threats and finished early.

The rest of the paper is structured as follows. In Section 
\ref{sec:treatments} we describe the treatments, and in Section 
\ref{sec:experiment} we provide the research questions and design of the case study. Section 
\ref{sec:res} presents the results of this work, while Section
\ref{sec:dis} discusses them. In Section 
\ref{sec:relatedwork} we position our work in the context of related work. Section 
\ref{sec:validity} discusses the limitations of this work, and Section 
\ref{sec:conclusion} gives the concluding remarks.
\section{Background}
\label{sec:treatments}
\looseness=-1
STRIDE is a methodology developed to help people identify the types of attacks their software systems are exposed to, especially because of design-level flaws.
The name itself is an acronym that stands for the threat categories of Spoofing, Tampering, Repudiation, Information Disclosure, Denial of Service and Elevation of Privilege.
For the definition of threat categories, we refer the reader to the documentation of STRIDE \cite{shostack2014threat}.
Our study compares two flavors of STRIDE.
The first technique is an example of a risk-last threat analysis technique, STRIDE-per-element \cite{shostack2014threat} (hereafter {\stride}).
The second is an example of a risk-first technique, the extended STRIDE \cite{tuma2017towards} (hereafter {\estride}).
The techniques differ in the model that is used for the analysis, and in the procedure of the analysis.

In {\stride} threats are first identified and then prioritized (therefore risk-last).
The first step is to create a graphical representation of the system architecture as a Data Flow Diagram (DFD).
A DFD represents how information moves around in a software-based system.
The diagram consists of processes (active entities), data flows (exchanged info), external entities (e.g., users or 3rd parties), data stores (e.g., file system) and trust boundaries.
The second step is a systematic exploration of the DFD graph to identify the threats.
For each element type, the methodology suggests looking into a subset of threat categories.
To this aim, STRIDE provides a table mapping element types to threat categories.
For instance, for external entities the analysts should look into Spoofing and Repudiation threats.
For each pair of element type and threat category, existing literature \cite{shostack2014threat} provides a catalog of example threats that can be used for inspiration by the analyst.

In {\estride} asset analysis is performed to identify important risks before threats are identified (therefore risk-first).
Thus, threat prioritization should not be necessary.
The first step is to create a regular DFD and extend it with security relevant information.
In addition to the DFD elements, the extended DFD (eDFD) notation explicitly models assets, their security objectives and priorities, asset sources and target elements, type of communication channels and domain assumptions.
This additional information needs to be manually extracted from the documentation and agreed upon.
The second step is a guided exploration of the eDFD to find security threats. 
The exploration is guided with rules that rely on the model extensions.
For instance, the assets with high-prioritized security objectives are traced in the eDFD.
Only security threats that directly threaten the high-priority objectives are considered.
For example, if confidentiality is of high priority, then information disclosure threats are considered.
Threats to elements with assumptions about existing security mechanisms are not considered.
These rules aim to reduce the manual effort.

\looseness=-1
Table \ref{tab:activities} shows the core activities of an analysis session using {\stride} and {\estride}.
\begin{table}
	\center
	\caption{Core activities of {\stride} and {\estride}.}
	\begin{tabular}{ p{0.13\columnwidth} p{0.45\columnwidth} cc}
		\toprule
		Step & Activity & {\stride} & {\estride} \\
		\midrule
		 Building & Drawing on the board & \Checkmark & \Checkmark \\
		 diagram & Architecture abstraction  & \Checkmark & \Checkmark \\
		  & Asset analysis  &  & \Checkmark \\
		  & Extending the diagram  &  & \Checkmark \\
		  & Focusing on critical architecture  &  & \Checkmark \\
		  & Scope discussion  & \Checkmark & \Checkmark \\
		\midrule
		 Analyzing & Attack scenario development & \Checkmark & \Checkmark \\
		 diagram & Domain discussion & \Checkmark & \Checkmark \\
		 & Threat feasibility discussion & \Checkmark & \Checkmark \\
		 & Threat consequence discussion & \Checkmark & \Checkmark \\
		 & Threat prioritization & \Checkmark &  \\
		 & Threat reduction & \Checkmark & \Checkmark \\
		\bottomrule
	\end{tabular}
\label{tab:activities}
\end{table}
Some activities are part of both techniques.
For instance, analysts have to abstract the architecture and make assumptions.
There are two differences between the core activities of {\stride} and those of {\estride}.
First, {\estride} carries out three additional activities (i.e., asset analysis, extending diagram, focusing on critical architecture) during the first step of the analysis.
Second, {\estride} does not prioritize threats after finding them (i.e., after diagram analysis).

\section{Design of Study}
\label{sec:experiment}
We conduct a case study where we compare the {\stride} analysis to the {\estride} analysis technique.
In what follows, we present the research questions, industrial case used in this study, and our participants.
We also describe the task given to the participants, the study execution, and the applied data collection methods and measures.

\subsection{Research questions}
This study answers research questions about the differences in the analysis outcomes (RQ1, RQ2) and procedure (RQ3) for the studied techniques.

\noindent
\textbf{RQ1.} \textit{What are the differences between a risk-last and a risk-first analysis technique in terms of productivity?}

Risk-last threat analysis prioritizes threats at the end of the analysis procedure.
In contrast, risk-first analysis aims to by-pass threat prioritization by analyzing the high risks first.
The purpose of the first research question is to understand whether by-passing threat prioritization in such a way helps to increase the amount of correctly identified threats per time unit.
In this respect, this research question is focused on investigating the differences in the analysis outcomes.

\noindent
\textbf{RQ2.} \textit{What are the differences between a risk-last and a risk-first analysis technique with respect to the timeliness and amount of discovered high-priority threats?}

\looseness=-1
In realistic circumstances, threat analysis sessions are pressed for time.
In this context, achieving complete coverage with a manual analysis is challenging.
Often threats are overlooked \cite{tuma2018two,scandariato2015descriptive} even if having strived for coverage during the analysis.
It seems reasonable to `knowingly' overlook less-important threats as compared to high-priority threats.
The purpose of the second research question is to investigate whether risk-first analysis leans important threats faster (and in a larger quantity) when compared to the risk-last analysis technique.
Similarly to RQ1, the second research question is focused on investigating the differences in the analysis outcomes.

\noindent
\textbf{RQ3.} \textit{What are the differences between a risk-last and a risk-first analysis technique with respect to the timeliness and amount of activities and activity patterns?}

Apart from the core activities (see Table \ref{tab:activities}), we observe important events and support activities taking place during the threat analysis session.
For instance, updating the diagram, or making an assumption.
Support activities include pointing at the board, taking a break, documenting, referring to case documentation, etc.
Due to the repetitive nature of manual threat analysis \cite{tuma2018two,scandariato2015descriptive}, these activities tend to re-occur.
We are interested to investigate which activities appear more often or sooner in both techniques.
In addition, we observe combinations of activities, or activity patterns to understand which technique better facilitates constructive thinking.
Therefore, the purpose of the third research question is to investigate the differences in the appearance of activities and activity patterns of the studied techniques.
Unlike RQ1 and RQ2, the third research question is focused on investigating the differences in the analysis procedure.

\subsection{Industrial case}
\looseness=-1
The industrial case used in this work is a reference architecture for an Electronic Control Unit (ECU) update using a mobile application in a vehicular setting. 
The reference architecture was provided by an automotive company.
This software is developed to reduce the vehicle downtime due to workshop visits.
The software is compatible with Android and Apple devices and is downloaded from the Application Store.
It is installed on a mobile device controlled by a driver or a technician.
A WiFi dongle is connected to the On-board diagnostics (OBD) port.
The mobile device connects to the WiFi dongle to update the ECUs.
Device owners (drivers and technicians) are able to use the mobile app to download ECU software updates from a remote software repository and install them on the ECUs of their vehicle.
The mobile app follows a procedure before updating the ECU software (e.g., checking the preconditions, such as device battery level, vehicle at rest, etc.).
Device owners are also able to use the mobile app to configure certain ECU parameters (depending on their level of authorization).
Similarly, a specific procedure is executed before updating the ECU parameters.
With respect to the confidentiality concerns we omit further details about the industrial case.

\subsection{Participants}
\label{subsec:participants}
\looseness=-1
The participants are industrial experts with experience in threat analysis.
We assembled two teams with 3 ({\stride}) and 4 members ({\estride}).
The {\estride} team had an additional member, a threat analysis trainee.
The authors made sure that the level of expertise was comparable in the two teams.
Each team member had an assigned role according to their expertise.
We include three roles in our case study: facilitator, security expert, and domain expert.
The trainee in {\estride} was assigned the role of a security expert, given the background of the participant.

\looseness=-1
\textit{Facilitators} were required to have an understanding of the STRIDE technique.
They were responsible for leading the session by following the prescribed procedure.
The facilitators also drew the diagrams and handled the documentation.

\textit{Security experts} were required to have technical knowledge about security mechanisms and attacks associated with critical infrastructure systems (e.g., banking and finance, telecommunications, transportation systems) \cite{hilburn2013software}.
They were responsible to extract relevant information from the domain expert and formulate attack scenarios.

\textit{Domain experts} were required to have a good over-all understanding of the system design and specialized knowledge about part(s) of the architecture \cite{hilburn2013software}.
They were responsible to contribute with relevant information for finding security threats.
Our domain experts have been employed by the organization for about ten years.

\subsection{Task}
The two teams were presented with the same task: perform a threat analysis on the industrial case using the prescribed technique.
The task was divided into two sub-tasks: participants were asked to (1) build a diagram based on the provided architectural documentation of the industrial case and (2) analyze the diagram according to the assigned technique.
The first sub-task required building a diagram (DFD for the {\stride} and eDFD for the {\estride} team).
The second sub-task required an analysis (systematic for the {\stride} and guided for the {\estride} team) of the created diagrams according to the steps described in Section \ref{sec:treatments}.
The analysis results had to be documented with a report and submitted electronically.
The report had to contain a list of identified threats, their location, attack scenarios, and estimated priority of the threat.
The participants were given a template for the report.
The purpose of the template is to simplify and standardize the analysis of the reports.
The template consisted of a unique ID of the diagram element, threat category, attack scenario, and threat priority.

\subsection{Execution}
\begin{figure*}
	\center
	\includegraphics[width=0.7\textwidth]{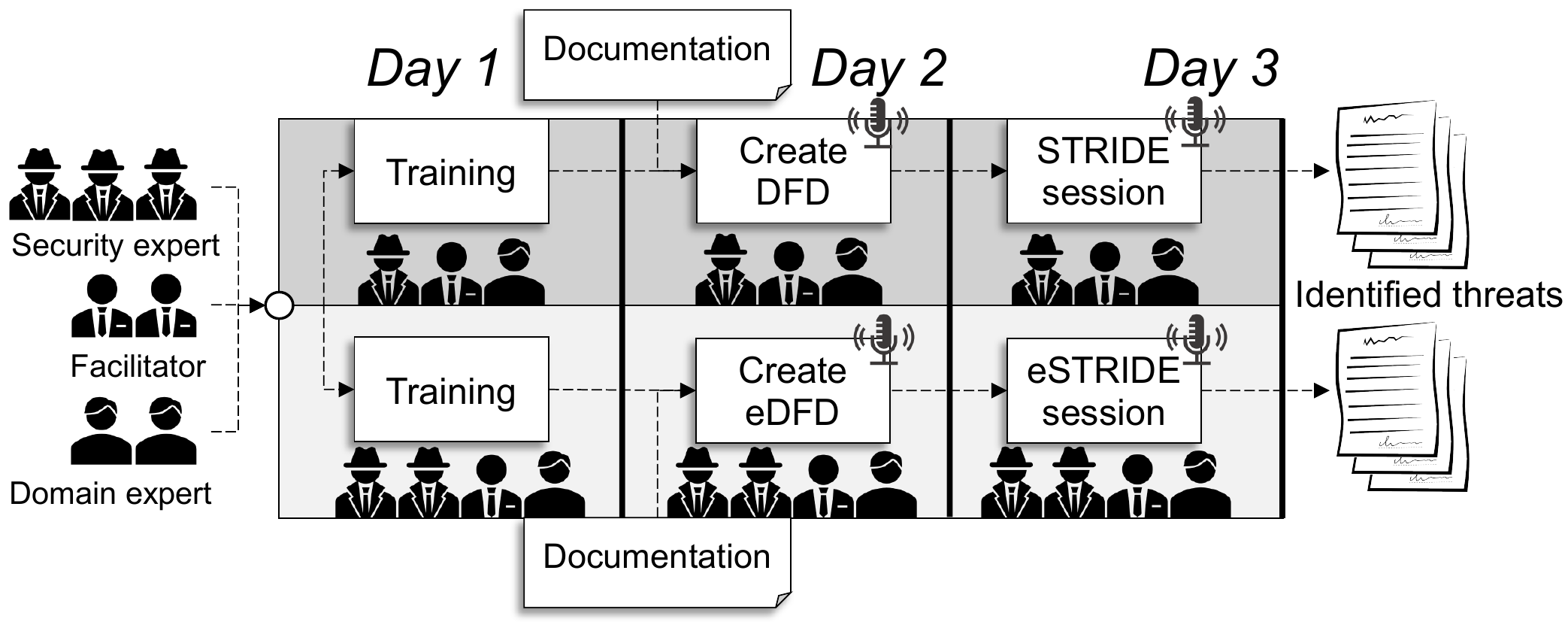}
	\caption{The execution of the case study.}
	\label{fig:execution}
\end{figure*}

Figure \ref{fig:execution} depicts the execution of the case study.
The workshops were split into three hour sessions taking place on three separate days during the same week.
The authors supervised all sessions that took place.
All the material (i.e., architectural documentation of the industrial case, training material, description of task, report template) was shared with the participants a week in advance.
Their were strongly encouraged to read all the material (about 20 pages) before the start of the case study.

\emph{Day 1: Training.}
On the first day, the teams were separated and trained (3h) to accomplish their task.
The training consisted of brushing up basic security concepts, building DFDs (or eDFDs), and performing STRIDE (or eSTRIDE).
In a separate training session, we warned both facilitators to monitor the progress and speed up the discussion, if necessary.

\emph{Day 2: Building diagram.}
On the second day, the teams were given printed copies of all the material and worked on their task.
The participants were allowed to continue with the second sub-task (diagram analysis) in case they finished the first sub-task.
With respect to complexity of the industrial case, we alloted only three hours per day to simulate realistic time constraints.
The authors observed and recorded the sessions, but did not participate in the discussion.

\emph{Day 3: Analyzing the diagram.}
On the third day, the teams were given the same printed material, and the diagram they had created the previous day.
The teams continued working on their task until they were either finished or ran out of time.
One participant of the {\stride} team was 30 minutes late on the third day of the case study.

\subsection{Data collection and measures}
We adopted a mixed methodology of qualitative and quantitative data analysis as suggested by Creswell et al. \cite{creswell2017research}.
First, we describe how the qualitative data (i.e., recordings of threat analysis sessions) were collected and what measures we adopted for the analysis.
Second, we describe the quantitative data (i.e., participant hand-ins) collection and measures.

\emph{Qualitative measures.}
We qualitatively analyzed the recorded sessions to answer RQ3.
The recordings were manually transcribed by the first author using software for qualitative analysis of interview data\footnote{\url{https://www.qsrinternational.com/nvivo/home}}.
The manual transcription process helped the researchers gain a deeper understanding of the recorded material.
After having a thorough understanding of the recordings, the first author \textit{coded} the transcriptions.
Coding is a technique for systematically marking chunks of transcriptions.
The analysis of code occurrences reveals trends and supports a qualitative analysis.
We used coding guidelines described by Wohlrab et al. \cite{wohlrab2016collaborative} to ensure the correctness of this step.
Table \ref{tab:eventssupport} depicts an iteratively developed hierarchy of codes.
\begin{table}
	\center
	\caption{Code hierarchy for observing analysis activities and events. Codes for events are marked by the $\boldsymbol{\dagger}$ symbol.}
	\begin{tabular}{ p{0.2\columnwidth}  p{0.7\columnwidth}}
		\toprule
		Level 0 & Level 1: Level 2 \\
		\midrule
		Building & Drawing on the board \\
		diagram & Architecture abstraction  \\
		& Asset analysis: Asset trace, Security concern, Concern value \\
		& Extending the diagram  \\
		& Focusing on critical architecture \\
		& Scope discussion \\
		& Making an assumption$\boldsymbol{\dagger}$ \\
		\midrule
		Analyzing & Attack scenario development \\
		diagram & Domain discussion \\
		& Threat feasibility: Importance, Attack scenario and surface  \\
		&  Threat consequence \\
		& Threat prioritization \\
		& Threat reduction \\
		& Using assumption$\boldsymbol{\dagger}$ \\
		& Updating diagram$\boldsymbol{\dagger}$ \\
		& High-priority threat found$\boldsymbol{\dagger}$ \\
		& Low or Medium-priority threat found$\boldsymbol{\dagger}$ \\
		\midrule
		Support & Pointing at board$\boldsymbol{\dagger}$ \\
		activities & Referring to task description$\boldsymbol{\dagger}$ \\
		& Referring to assumptions$\boldsymbol{\dagger}$ \\
		& Referring to case document$\boldsymbol{\dagger}$ \\
		& Referring to training material$\boldsymbol{\dagger}$ \\
		& Break$\boldsymbol{\dagger}$ \\
		& Unsure \\
		& Documenting \\
		\midrule
		Detour & Chatting \\
		& Difference in opinion \\
		& Terminology \\
		\bottomrule
	\end{tabular}
	\label{tab:eventssupport}
\end{table}
We coded activities related to \textit{diagram building}, \textit{analysis of diagram}, \textit{support activities}, and \textit{detour} activities.
We used the same codes for coding activities in both teams.
With respect to diagram building, we coded architecture abstraction, assets analysis, making assumptions, scope discussion, etc.
Regarding diagram analysis, we coded attack scenario development, threat feasibility discussion, threat consequence, and the like. 
After assessing the participants' hand-ins (described in \textit{Quantitative measures}), we also marked important events (marked with $\dagger$ in Table \ref{tab:eventssupport}) in the transcriptions.
For instance, a correct discovery of a high-priority threat.
Regarding detour activities, we coded terminology discussion, difference in opinion, and chatting.
Finally, we coded supporting activities (e.g., drawing on the board).
These codes were manually inserted in the transcriptions.

\emph{Quantitative measures.}
We quantitatively analyze the reports handed in by our participants to answer RQ1 and RQ2.
The first author assessed the participant hand-ins and assigned priorities to the reported threats (high, medium, low).
The hand-ins included pictures of the created diagrams, list of assumptions, and a list of identified threats (documented according to the provided template).
This assessment was validated by the industrial experts. 
This measurement allows us to identify differences between the two groups across the priority classes.

This work refers to a true positive $(TP)$ as a correctly identified threat. 
This means that: 
(a) the participants found the threat in the correct diagram location,
(b) the participants found a realistic attack scenario for the security threat or the participants found a security vulnerability, and 
(c) the threat is correct with respect to the assumptions the participants made.

We refer to a false positive $(FP)$ as an incorrectly identified security threat. 
This means that the participants found the security threat in the wrong location or the threat is not correct with respect to the assumptions the participants have made.

In addition, we marked security threats that might be correct, but could not be assessed as correct due to insufficient information $(II)$. 
The reported threats were marked with insufficient information if 
(a) the participants found the security threat in the right location of the diagram, and 
(b) the threat is correct with respect to the assumptions the participants have made, but 
(c) the participants neither found a realistic attack scenario nor identified a security vulnerability. 

Precision ($P=\frac{TP}{TP+FP}$) was measured as the ratio between correctly identified threats and all identified threats.
Productivity ($Prod=\frac{TP}{h}$) was measured as the amount of correctly identified threats per hour.

In addition, we quantitatively analyzed the transcribed recordings to answer RQ3.
As mentioned before (see \emph{Qualitative measures}), coding transcriptions enabled us to track the exact location (e.g., index in the text where the activity starts) of a particular activity in the transcript.
We analyze the recordings by observing activity code locations (and the distance between them) in the transcriptions.
There is an element of time in the distances between activities.
For instance, two activities can be close in the transcription, but in reality, there may have been a pause in the conversation.
Time is not considered explicitly in our distance measure.
We do not use timestamps to measure the distance because, even is inserted often, timestamps can only provide a coarse-grained estimation of the distance between activities.
Instead, we normalize the average distance with the length of the transcription.
In this way, we implicitly consider time in our distance measure.
Therefore, the spatial distance (i.e., amount of characters in text) between activities works as a proxy of the temporal distance.
The average distance was measured as the average number of characters separating the starting indexes of the code occurrences.
Let $A$ be a code occurring $n$ times in transcription $TR$, and let $B$ represent a different code occurring $m$ times in transcription $TR$ $(m,n \in \mathbb{N}, \ and \ m,n > 0)$.
The number of characters contained in $TR$ is denoted by $len(TR)$.
Let also $o$ be some occurrence in transcription $TR$ and let $indexof(c)$ be the index of the first character of the occurrence in $TR$.
Then, the average distance $dist$ between code $A$ and $B$ was obtained as follows:
\begin{equation}
dist = \frac{\sum_{i,j=0}^{m,n} (| \ indexof(o^{A}_i) \ - \ indexof(o^{B}_j)  |) \times 100}{i \ \times j \ \times \ len(TR)}
\end{equation}
In other words, $dist$ between $A$ and $B$ shows the average percentage of the transcription chuck separating the code pair.
For instance in {\stride}, the code threat consequence and threat prioritization are on average separated by 2.23 \% of the entire transcription (see Table \ref{tab:dist}).

\section{Results}
\label{sec:res}
This section reports the results of the quantitative and qualitative analysis of the collected data.

\subsection{RQ1: Productivity}
Table \ref{tab:qualitative} shows the results obtained from assessing the hand-ins of the participants. 
\begin{table}
	\center
	\caption{A quantitative assessment of the hand-ins.}
	\label{tab:qualitative}
	\begin{tabular}{p{3.5cm} c c c c}
		\toprule
		& & {\stride} & {\estride} & Common\\
		\midrule
		Correct threats $(TP)$ & H & 4 & 8 &  4\\
		& M & 2 & 1 & 0\\
		& L & 6 & 4 & 2\\
		\cmidrule(lr){3-5}
		Total &  & 12 & 13 & 6\\
		\midrule
		Incorrect threats $(FP)$ & & 0 & 0 & - \\
		Insufficient info $(II)$ & &  15 & 0 & -\\
		\midrule
		Precision $TP/(TP+FP)$ & & 1 & 1 & \\
		Productivity $TP/h$ & & 3 & 2.6 & \\
		\bottomrule
	\end{tabular}
\end{table}
\looseness=-1
Overall, the {\stride} team documented 27 threats, whereas the {\estride} team documented only 13 threats. 
This is not surprising, considering that {\estride} guides the analysis towards threat reduction.
Neither {\stride} not {\estride} team documented incorrect threats (FPs).
The amount of TPs is not different for the compared techniques ({\stride}: 12, {\estride}: 13).
Further, no big differences were observed in the productivity of the compared techniques ({\stride}: 3 vs {\estride}: 2.6).

All the threats that were marked with $II$ were thoroughly discussed with the co-authors from the automotive industry.
Note that the authors tried to be conservative with discarding threats.
As per the measures mentioned before (Section \ref{sec:experiment}), the authors only marked threats with $II$ if the participants failed to identify an attack scenario or a security vulnerability.
To verify this, all the collected data were examined (i.e., hand-ins, and recordings).
We discuss this matter further in Section \ref{sec:dis}.
About half of the documented threats (15) from the {\stride} team were marked with insufficient information $(II)$.
A deeper analysis of the hand-ins revealed differences in the level of detail each identified threat was described.
For instance, the {\estride} team documented an information disclosure threat to the mobile app as:
\begin{myquote}{0.1in}
	\textit{``Data collection of \dots, which can be used to access \dots without being connected to \dots ''}
\end{myquote}
On the other hand, the {\stride} team documented the same threat as:
\begin{myquote}{0.1in}
	\textit{``We can lose the credentials. The phone can be hacked.''}
\end{myquote}
Further, the recordings showed that the {\estride} team developed and discussed a feasible attack scenario where such data collection could occur.
On the other hand, the {\stride} team agreed that the phone can be hacked, yet did not discuss how this can occur.
For this reason, such a threat was marked with $II$.
Note that, the {\stride} team did not perform a careless analysis, despite having a very quick pace. 
 
\subsection{RQ2. Discovering high-priority threats}
\begin{figure*}
	\center
	\includegraphics[width=0.9\textwidth]{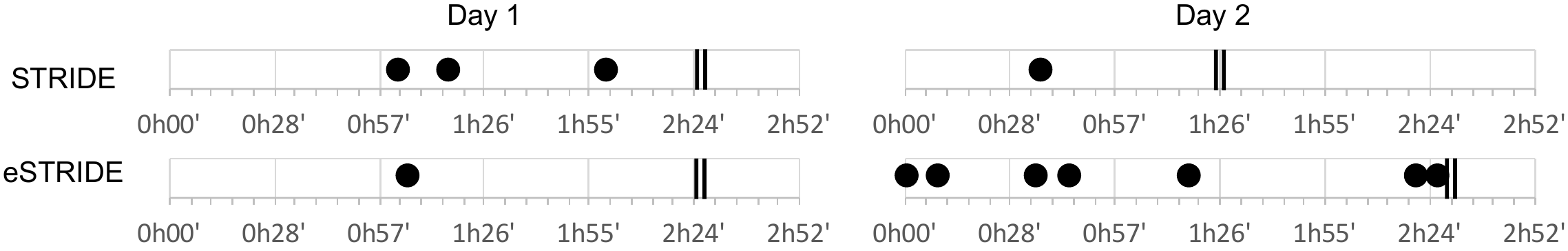}
	\caption{High-priority threats discovered by the {\stride} (top) and {\estride} (bottom) team.}
	\label{fig:htp}
\end{figure*}

Figure \ref{fig:htp} depicts when the teams discovered high-priority threats.
The {\stride} team discovered 4 high-priority threats, three of which they discovered already during the first day. 
The last high-priority threat was discovered by the {\stride} team about 40 minutes into the second day.
Note that, the {\stride} team started analyzing the diagram one hour into the first day.
The {\estride} team discovered 8 high-priority threats, most of them during the second day. 
Yet, they discovered one high-priority threat already during the first day 
while discussing security objectives of assets and their values.
The {\estride} team started analyzing the diagram on the second day.
Compared to the {\stride} team, the {\estride} team did not find high-priority threats faster.

\subsection{RQ3. Focus of activities and activity patterns}
We investigate similarities and differences in activity focus, time-lines of activities, and activity patterns to answer RQ3.
First, we report on the focus of activities by presenting the code coverage of the transcriptions.
Second, we present a detailed time-line of activities.
Finally, we report on the activity pairs by presenting their average distance in the transcriptions.

\begin{figure}
	\center
	\includegraphics[width=0.9\linewidth]{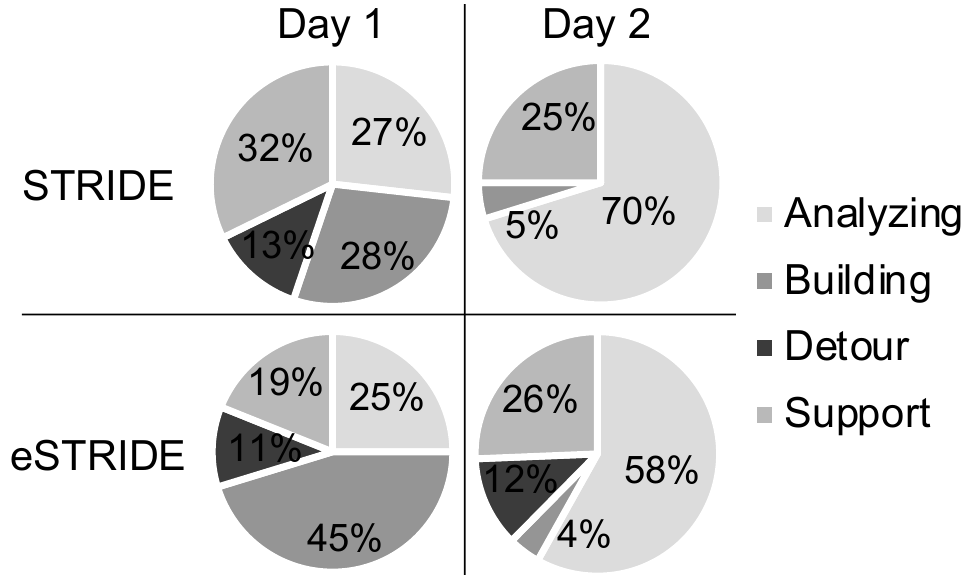}
	\caption{Percentage of coding references for {\stride} (top) and {\estride} (bottom).}
	\label{fig:day12pie}
\end{figure}

\textbf{Focus of activities.}
Figure \ref{fig:day12pie} shows the percentage of coding references for high-level activities extracted from NViVo.
During the first day the {\stride} team did not focus on one particular activity due to having started with the second step of the analysis (see Section \ref{sec:treatments}) quite early.
Diagram analysis covered 27\% of the recorded session (mainly discussing the domain and developing attack scenarios).
Building the diagram covered 28\% of the transcription. 
The team detoured often from the prescribed procedure (13\%). 
Finally, the {\stride} team was involved in more support activities during the first day (32\%). 
The {\estride} team did not start with the second step of the analysis.
Rather, they analyzed assets when extending the diagram (sec Section \ref{sec:treatments}).
Similarly, the {\estride} team detoured often during the first day (11\%).
In contrast to {\stride}, the support activities covered only 19\% of the transcription during the first day.
Further, the {\estride} team focused on diagram building (45\%). 

During the second day, the {\stride} team focused on diagram analysis (70\%).
Support activities covered a quarter of the {\stride} transcription (25\%).
The team made minor changes to the diagram (5\%) but did not detour from the analysis.
Similarly, during the second day the {\estride} team focused on diagram analysis (58\%) and support activities (26\%) and made minor changes to the diagram (4\%).
In contrast, the {\estride} team detoured (12\%) from the analysis procedure during the second day.

\textbf{Time-line of activities.}
We analyzed time-lines of code occurrences to answer RQ3.
\begin{figure*}
	\center
	\includegraphics[width=1.0\textwidth]{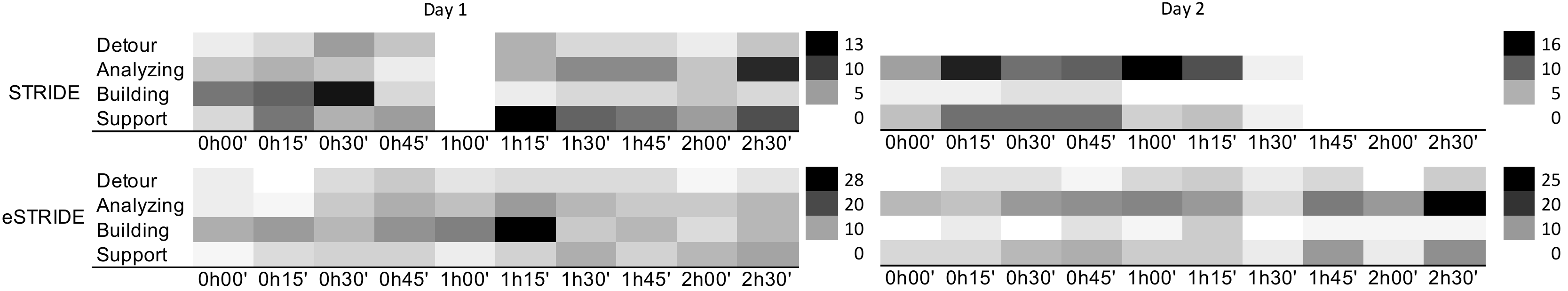}
	\caption{Time-lines of code occurrences for the {\stride} (top) and {\estride} team (bottom).}
	\label{fig:timelinesday1day2}
\end{figure*}
\looseness=-1
Figure \ref{fig:timelinesday1day2} depicts the time-lines of activities for both teams.
The time-lines show the aggregated number of code occurrences per ten-minute time frame.
Note that, the {\stride} transcription is almost half the size of the {\estride} transcription (90,612 vs 151,907 characters).
This explains the different proportion of code occurrences in the time-lines.
In what follows, we discuss the similarities and differences in activities during the first day and the second day.

\subsubsection*{Similarities (Day 1)}
\looseness=-1
In the first 15 minutes both teams focused on building the diagram.
In particular, both teams focused on abstracting the architecture, discussing the domain, discussing the scope, and drawing on the board.
Other support activities in this time-window include referring to the case documentation.
In the span of the entire session, both teams sometimes detoured from the instructed analysis procedure.
The detours during the first day are fairly evenly distributed across teams.
Both teams made the assumptions during the first day, and made one last assumption about one hour into the second day.
\subsubsection*{Differences (Day 1)}
\looseness=-1
About an hour into the first day, both teams focused on support activities (i.e., referring to case documentation).
The {\stride} team finished building the diagram after about an hour.
They read parts of the case documentation aloud to validate the diagram before they started to analyze it.
On the other hand, the {\estride} team started extending the diagram with domain assumptions after about an hour.
They verified each assumption by reading the case documentation aloud.
The {\estride} team started looking for threats only on the second day.
In contrast to {\stride}, the {\estride} team made over-all less assumptions and documented them early-on.
The {\stride} team agreed upon some assumptions but did not document them.

\subsubsection*{Similarities (Day 2)}
\looseness=-1
As instructed, both teams performed activities related to diagram analysis which are accompanied by support activities (mainly, documenting threats).
Roughly speaking, the participants alternated between analyzing the diagram and documenting threats.
This pattern is more apparent in Figure \ref{fig:timelinesday1day2} ({\estride} team), as the {\stride} team was very quick in documenting threats.
Both teams had a strong focus on diagram analysis in two time-frames ({\stride} 01:10:00-01:15:00 and 01:20:00-01:25:00, {\estride} 01:20:00-01:25:00 and 02:00:00-02:05:00).
In all four cases, the teams managed to thoroughly analyze one threat in a span of five minutes.
This entailed developing attack scenario, using an assumption, discussing threat consequence, determining feasibility, and finding a correct threat.
\subsubsection*{Differences (Day 2)}
\looseness=-1
Compared to the first day, both teams detoured less from the instructed analysis procedure.
In particular, the {\stride} team did not detour at all.
In fact, the {\stride} team finished about one hour earlier.
Compared to {\estride}, the {\stride} team focused less on feasibility analysis during the second day 
and attack scenario development.
The {\stride} team often updated their diagram during the second day.
Specifically, the team merged data flows and removed one external entity and three data stores.

In summary, during the first day the {\estride} team spent more time building the diagram and during the second day, the {\stride} team did not detour from the analysis procedure.
We further discuss this in Section \ref{sec:dis}.

\textbf{Distance between activity pairs.}
We calculated average distances between all activity pairs for both teams.
\begin{table}
\center
\caption{The differences between activity distances in {\stride} and {\estride}. In case of a small difference, activity codes A and B have a similar average distance in both teams.}
\label{tab:dist}
\begin{tabular}{ p{0.5\columnwidth} p{0.1\columnwidth} p{0.1\columnwidth} p{0.1\columnwidth}}
\toprule
Code A  \& Code B  & {\stride} & {\estride} & $\Delta$ dist \\
\midrule
Threat reduction \& Ref. to assumptions & close & close & 0.10 \\ 
Terminology \& Domain discussion  & close & close & 1.70 \\
High-priority threat found \& Attack scenario or vulnerability & close & close & 1.84 \\ 
\midrule
Asset analysis \& Updating diagram  & far & close& 29.0 \\ 
Ref. to training material \& Unsure  & close & far & 38.38\\ 
Scope discussion \& Updating diagram  & far & close & 38.24\\ 
\bottomrule
\end{tabular}
\end{table}
\begin{figure}
	\center
	\includegraphics[width=0.5\textwidth]{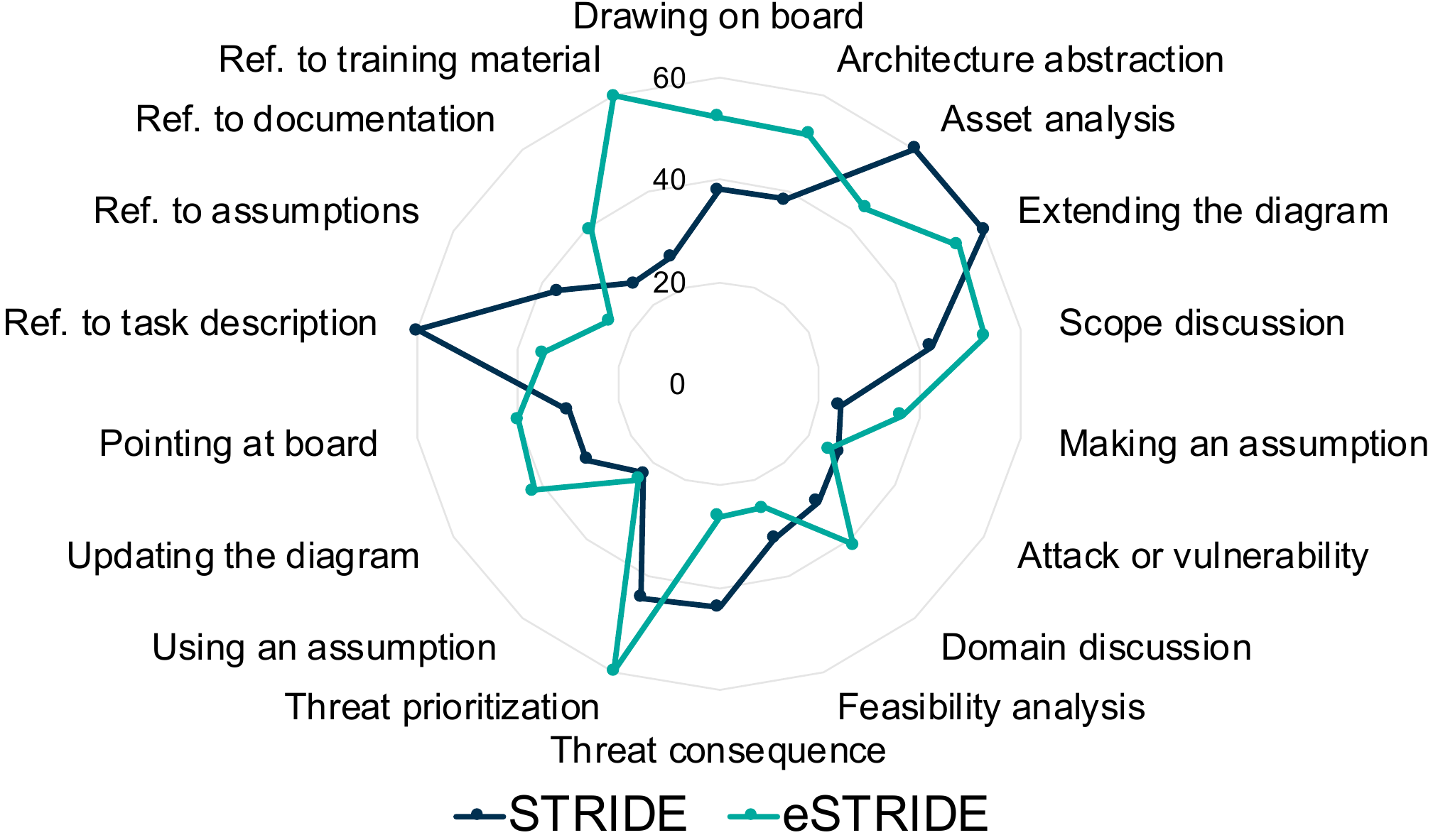}
	\caption{The average distance between finding a high-priority threat and other activities for both teams. The smaller the distance, the closer the activity to finding a high-priority threat.}
	\label{fig:spiderhtp}
\end{figure}
Table \ref{tab:dist} depicts the difference ($\Delta$) between the activity pair distances of {\stride} and {\estride}.
We omit some entries from Table \ref{tab:dist} due to space limitations.
For example, in both teams drawing on the board was close to architecture abstraction, as expected, and therefore omitted.
We discuss the similarities (smallest differences in the top part of Table \ref{tab:dist}) and differences (largest differences in the bottom part) between teams.
In addition, we analyzed the distances between single activities in relation to all other activities for both teams.
For instance, Figure \ref{fig:spiderhtp} shows the average distance between finding a high-priority threat and other activity codes for both teams.
\subsubsection*{Similarities}
\looseness=-1
Both teams referred to their assumptions during threat reduction to make sure the reductions do not lead to overlooked threats ($\Delta dist = 0.10$).
When the teams referred to assumptions, they read the assumption out loud.
This is interesting, as the experimenters did not instruct them to perform this step.
In addition, both teams engaged in a domain discussion while clarifying the terminology.
Finally, both teams found high-priority threats while developing attack scenarios or identifying vulnerabilities.

Figure \ref{fig:spiderhtp} shows, that the average distance between using assumptions and finding high-priority threats is small in the transcriptions of both teams.
The teams used assumptions to justify their reasoning for a threat or vulnerability existence.
The {\estride} team also read the assumptions out loud.
Therefore, the average distance between referring to assumptions and a finding high-priority threat is small in the {\estride} transcriptions.
%

\subsubsection*{Differences}
\looseness=-1
In contrast to {\stride}, the {\estride} team performed an asset analysis and iteratively updated the diagram with the extra security information ($\Delta dist = 29.0$).
In addition, the {\estride} team discussed the scope of the analysis while updating the diagram.
For instance, they discussed which parts of the system can be left out of the analysis (assumed as trusted).
This was not discussed at length in the {\stride} team.
During the first day, {\stride} team referred to the training material when unsure.

Compared to {\stride}, the average distance between finding important threats and discussing threat feasibility (and consequence) is smaller in the {\estride} transcription (see Figure \ref{fig:spiderhtp}).
Further, the {\estride} team found the first high-priority threat when analyzing the assets and extending the diagram in the first day.
Compared to {\estride}, the average distance between finding important threats and referring to training and case documentation is smaller in the {\stride} transcription.
In fact, the {\stride} team relied more on the support material, whereas the {\estride} team relied more on the domain expert.
This may be due to factors of team dynamics, rather then the differences in the techniques.
Finally, the {\stride} team made several assumptions during diagram analysis, therefore the average distance between making assumptions are finding important threats is smaller, compared to the {\estride} transcription.

In summary, for both teams assumptions played an important role in finding high-priority threats and in reducing threats.
In addition, developing attack scenarios and discussing threat feasibility supported finding high-priority threats (more so in the {\estride} team).
However, our analysis indicates that differences in activity patterns might depend on factors related to team dynamics rather then the differences in the techniques.

\section{Discussion}
\label{sec:dis}
In this section we discuss the results and answer the research questions.

\subsection{RQ1: Productivity}
Our assessments show that most of the $II$ marked threats (13 out of 15) of {\stride} were of low-priority.
Yet, half of these threats (8 our of 15) were still discussed in the {\estride} team.
In such cases, the {\estride} team identified a possible attack scenario, but found it infeasible.
In one example, a correctly identified a high-priority threat by the {\estride} team was marked with $II$ in the {\stride} team.
The rest of these threats (6 out of 15) were skipped in the {\estride} due to the {\estride} reductions.

Six security threats were correctly discovered by both teams (fourth column in Table \ref{tab:qualitative}).
Four out of those common threats were high-priority threats, and two were of low-priority.
The {\stride} team discovered 6 threats that were not discovered by the {\estride} team (2 medium and 4 low-priority).
In such cases, the {\estride} team either skipped this location due to {\estride} reductions (2 low, 1 medium) or agreed that the attack is not feasible (2 low, 1 medium), and documented:
\begin{myquote}{0.1in}
	\textit{``No interesting attack scenario.''}
\end{myquote}
The {\estride} team discovered 5 threats that went unnoticed by the {\stride} team.
In contrast, these threats were of high (4) and medium priority (1).
In these cases, the {\stride} team could not find any vulnerability of attack, and documented, for instance:
\begin{myquote}{0.1in}
	\textit{``Repudiation is not a problem.''}
\end{myquote}
A possible explanation is that the {\stride} team may not have discussed threat feasibility enough to find feasible attack scenarios or that they were simply overlooked.

\smallbreak
\noindent
\fbox{\begin{minipage}{0.96\columnwidth}
We investigated productivity to answer \textbf{RQ1}.
We did not observe a difference in the productivity of the teams.
However, the {\estride} team has found more threats of high-priority (4) which were overlooked by the {\stride} team. 
In contrast, the {\stride} team discovered more threats of low-priority (4) and finished early. 
This explains the slightly better productivity in Table \ref{tab:qualitative}. 
\end{minipage}}

\subsection{RQ2: Discovering high-priority threats}
\looseness=-1
Both teams started exploring the diagram from external entities, where some high-priority threats we located.
Therefore, the first few high-priority threats were correctly discovered by both teams at the very beginning of diagram analysis.
The {\stride} team chose to continue the exploration with the processes (this was completed at the end of the first day).
On the second day, the {\stride} team systematically explored all the data flows.
In contrast, the {\estride} team continued the exploration with important assets in an end-to-end fashion.
Possibly, this may have helped in finding more high-priority threats.
Yet, extending the diagram with security-relevant information required more effort in the beginning of the analysis.

\smallbreak
\noindent
\fbox{\begin{minipage}{0.96\columnwidth}
		Concerning \textbf{RQ2}, we found that the {\estride} team found double the amount of high-priority threats (8) compared to the {\stride} team (4).
		Further, all high-priority threats that were discovered by the {\stride} team were also discovered by the {\estride} team.
		In the settings of this case study, the {\estride} team was more complete with respect to finding high-priority threats.
		Yet, we do not have enough evidence to conclude that {\estride} can identify high-priority threats sooner.
\end{minipage}}

\subsection{RQ3: Focus of activities and activity patterns}
Sometimes, the teams discussed terminology which lead to a difference in opinion about threat categories (more so in the {\stride} team).
In particular the spoofing category was often discussed (in relation to tampering and repudiation).
Detours can be minimized by the facilitator steering the discussion.
The {\stride} team often referred to the material to reach consensus, instead.
Possibly, this motivated the team to stay closer to the instructed procedure on the second day.
On the second day, the domain expert was 30 minutes late, during which the team started working with a high pace.
In the {\estride} transcription, the distance between detours and threat feasibility analysis was small.
Therefore, feasibility analysis may have slowed down the analysis.
Discussing threat feasibility often leads to estimating the probability of threat occurrence, which is difficult and can lead to `analysis paralysis'.
Yet, the facilitator of {\estride} often steered the discussion back on track. 

\looseness=-1
\smallbreak
\noindent
\fbox{\begin{minipage}{0.96\columnwidth}
		Regarding \textbf{RQ3}, we found that the {\estride} team took longer to build the diagram, but managed to capture more high-priority threats.
		Compared to {\stride}, the {\estride} team discussed threat feasibility in more detail and developed detailed attack scenarios.
		On the other hand, the {\stride} team performed the analysis with a quick pace and spent less time on building the diagram.
		For what concerns the activity patterns, we found that both teams were careful when making threat reductions, backing those decisions by referring to assumptions.
		In addition, assumptions were used by both teams to justify the existence of threats (in particular high-priority).
		Our analysis indicates that differences in activity patterns might depend on factors related to team dynamics.
\end{minipage}}

\section{Related Work}
\label{sec:relatedwork}

Recently, Stevens et al. \cite{stevens2018CoG} conducted a case study investigating the efficacy of threat analysis in an enterprise setting.
The authors develop qualitative measures to determine the efficacy of the Center of Gravity (CoG) technique.
The CoG originated in the 19th century as a military strategy and is by nature a risk-first technique.
The authors design a six-step protocol (including surveys and classroom sessions) and involve 25 practitioners in the study.
Similarly to this study, they report a very high accuracy of the results handed-in by industrial practitioners.
In addition, they provide empirical evidence for a perceived usefulness of threat analysis even after 30 and 120 days, which is very promising.
Our study is novel in that it investigates the timeliness of high-priority threats, and the activity focus of a risk-first and a risk-last technique.

\looseness=-1
McGraw conducted a study including 95 companies \cite{BSIMM}. 
The study reports on the security practices that are in place in these companies.
The BSIMM model does not mention STRIDE per se, rather it highlights the importance of threat analysis.
Microsoft has not published evidence of the effectiveness of the STRIDE-per-element technique \cite{shostack2014threat}.
Similarly, eSTRIDE (coupled with eDFD) \cite{tuma2017towards} is a recently proposed technique, evaluated solely on the basis of an illustration.

Tuma et al. \cite{tuma2018two} conducted a controlled experiment comparing the two STRIDE variants, STRIDE-per-element and STRIDE-per-interaction. 
Similarly to this work, their study quantitatively measures the precision, and productivity of both variants.
Their study concludes that there is no statistically significant differences in precision, recall, and productivity of the two STRIDE variants. 
Yet, the authors speculate that enlarging the analysis scope from one (or two) elements to an end-to-end scenario might have an effect on performance.
Their findings are based on quantitative measures, while we adopted a mixed methodology, including a qualitative analysis of recorded sessions.

\looseness=-1
Scandariato et al.~\cite{scandariato2015descriptive} have analyzed STRIDE-per-element and evaluated the productivity, precision, and recall of the technique in an academic setting.
The purpose of their descriptive study was to provide an evidence-based evaluation of the effectiveness of STRIDE.
Our study, on the other hand, provides a comparative evaluation (by means of a controlled experiment) of STRIDE-per-element and the recently proposed eSTRIDE.

Labunets et al~\cite{labunets2013experimental} have performed an empirical comparison of two risk-oriented threat analysis techniques by means of a controlled experiment with students.
The aim of the study was to compare the effectiveness and perception of a visual technique with a textual technique.
The main finding of this study shows that the visual method is more effective for identifying threats than the textual one, while the textual method is slightly more effective for eliciting security requirements.

Existing literature reports on different measures, such as perception of techniques compared to misuse cases (MUC).
The work of Karpati, Sindre, Opdahl, and others provide experimental comparisons of several techniques.
Opdahl et al.~\cite{opdahl2009experimental} measure the effectiveness, coverage and the perception of the techniques.
Karpati et al.~\cite{karpati2011experimental} present an experimental evaluation of MUC Map diagrams focusing on identification of not only vulnerabilities but also mitigations.
Finally, Karpati et al.~\cite{karpati2012comparing} have experimentally compared MUCs with mal-activity diagrams in terms of efficiency.

\section{Threats to Validity}
\label{sec:validity}
We briefly discuss the main limitations of this study. 
With respect to the \textit{internal threats} to validity, we mention the confounding factors that may have influenced the results. 
The most important confounding factor is team dynamics. 
The performance of a team might depend on how well the participants work together. 
High performing teams have typically been working together for longer period of time \cite{losada1999complex}. 
The researchers were involved in selecting the members of the teams thus introducing potential bias in the study. 
Further, one team ({\estride}) consisted of an additional member (trainee). 
In light of the circumstances and limited resources, such a selection was necessary.
To mitigate this treat, the authors took careful notes during the execution of the study.
We believe that this did not significantly impact the performance of the {\estride} team.
Some participants may have worked together before the case study, while others might have collaborated less in the past. 
Another important confounding factor is different background knowledge across teams. 
We control for this factor by dedicating a whole workshop (3 hours) to training the participants.
We also mention the risk of subjectivity in the data analysis and the authorship of one of the techniques ({\estride}). 
The distance between activities is measured with a spatial measure, as compared to a temporal measure, thus this measure is more fine grained.
To counter the possible effect of time, we have normalized the measured distances.
Finally, the participants of the {\estride} team could have understood that this is a new technique and felt more motivated.
This could have introduced participant bias.

\looseness=-1
With respect to the \textit{external threats} to validity, we consider the threat to generalizability of the results.
The study was not repeated with other participants or with other industrial cases.
In addition, the number of participants was small and is thus hardly a representative sample of the population.
To mitigate this threat, we focused on conducting the case study in a realistic setting and relied more on the qualitative analysis results.

\section{Conclusion}
\label{sec:conclusion}
This study investigates the benefits and shortcomings of performing a risk-first ({\estride} \cite{tuma2017towards}) compared to risk-last ({\stride} \cite{shostack2014threat}) threat analysis in an industrial setting.
We conduct a case study with industrial experts.
In this setting, we gather empirical evidence about the performance and execution of the two techniques.  
The contributions of this work are three-fold: 
(i) a quantitative comparison of performance, 
(ii) a quantitative and qualitative comparison of execution, and
(iii) a comparative discussion of the benefits and shortcomings of the two techniques.
This study found no differences in the productivity and timeliness of discovering high-priority security threats.
But, we show differences in analysis execution.
In particular, the team using the risk-first technique found twice as many high-priority threats, developed detailed attack scenarios, and discussed threat feasibility in detail.
On the other hand, the team using the risk-last technique found more medium and low-priority threats and finished early.
We plan to replicate the case study with more participants and different industrial cases.
An interesting future direction is also conducting a longitudinal study to understand whether {\estride}'s benefits (prioritizing the discovery of high-priority threats) out-weight the limitations (required effort to build eDFDs and sacrificed coverage of low-prioritized threats).

\bibliographystyle{ACM-Reference-Format}
\bibliography{lit} 

\end{document}